\documentclass[twocolumn,aps]{revtex4} 

\usepackage{epsf}

\begin{document}

\title{Coupled Cluster Treatment Of An Interpolating 
 Triangle/Kagom\'e Antiferromagnet}

\author{D.J.J. Farnell, R. F. Bishop, and K.A. Gernoth}

\affiliation{Department of Physics, University of Manchester Institute of
  Science and Technology (UMIST), P O Box 88, Manchester M60 1QD, United 
  Kingdom}

\date{\today}

\begin{abstract}
The coupled cluster method (CCM) is applied
to a spin-half model at zero temperature which interpolates 
between a triangular lattice antiferromagnet (TAF) and a 
Kagom\'e lattice antiferromagnet (KAF).
The strength of the bonds which connect Kagom\'e lattice sites 
is $J$, and the strength of the bonds which link the 
non-Kagom\'e lattice sites to the Kagom\'e lattice sites
on an underlying triangular lattice is $J'$.
Our results are found to be highly converged, and our 
best estimate for the ground-state energy per spin 
for the spin-half KAF ($J'=0$) of 
$-$0.4252 constitutes one of the most accurate 
results yet found for this model. 
The amount of classical ordering on the Kagom\'e lattice sites 
is also considered, and it is seen that this parameter goes to 
zero for values of $J'$ very close the KAF point. 
Further evidence is also presented for CCM critical points 
which reinforce the conjecture that there is a phase near 
to the KAF point which is much different to that near to 
the TAF point ($J=J'$).

\begin{flushleft}
\end{flushleft}
\end{abstract}

\maketitle

Our knowledge of 
the zero-temperature properties of lattice quantum spin systems 
has been enhanced by the existence of 
exact solutions, mostly for $s=1/2$ one-dimensional systems, and 
by approximate calculations for higher quantum spin number and 
higher spatial dimensionality. Of particular
note have been the density matrix renormalisation group (DMRG)
calculations \cite{DMRG1} for one-dimensional (1D) 
and quasi-1D spin systems, although the DMRG has, as yet, not been so 
conclusively applied to systems of higher spatial dimensionality. 
Similarly, quantum Monte Carlo (QMC) calculations 
\cite{qmc3,qmc4} at zero temperature are limited by 
the existence of the infamous sign problem, which in turn is often a 
consequence of frustration for lattice quantum spin systems. We 
note that for non-frustrated systems one can often determine 
a ``sign rule'' \cite{sign_rules1,sign_rules2} which completely 
circumvents the minus-sign problem.

A good example of a spin system for which, as yet, no sign
rule has been proven is the spin-half triangular lattice Heisenberg 
antiferromagnet (TAF). The fixed-node quantum Monte Carlo (FNQMC) method 
\cite{taf1} has however been applied to this system with some success,
although the results constitute only a variational upper bound
for the energy. Other approximate methods \cite{taf2,taf3,taf4,taf5}
have also been successfully applied to the spin-half TAF, and most,  
but not all, such treatments predict that about 50$\%$ of the classical 
N\'eel-like ordering on the three equivalent sublattices remains in the
quantum case. In particular, series expansion results \cite{taf2}
give a value for the ground-state energy of $E_g/N$=$-$0.551,
although the corresponding value for the amount of remaining classical 
order of about 20$\%$ is almost certainly too low.
This spin-half TAF model therefore constitutes a very 
challenging problem for such approximate methods. However, 
the spin-half Kagom\'e lattice Heisenberg 
antiferromagnet (KAF) poses an even more difficult problem,
because, like the TAF, not only is it highly frustrated and 
no exactly provable ``sign rule'' exists, but also the classical
ground state is infinitely degenerate. Careful finite-sized 
calculations \cite{kaf1,kaf2,kaf3} have however been performed for 
the quantum spin-half KAF, and these results indicate that 
none of the classical N\'eel-like ordering seen in the TAF 
remains for the quantum KAF model. The best estimate 
for the ground-state energy of the KAF via finite-sized
calculations \cite{kaf3} stands at $E_g/N$=$-$0.43.
Furthermore, series expansion results \cite{taf2} indicate  
that the ground-state of the KAF is disordered. Indeed,
a variational calculation \cite{kaf4} which utilised a dimerised 
basis also found that the ground state of the KAF 
is some sort of spin liquid.

\begin{figure}
\epsfxsize=8cm
\centerline{\epsffile{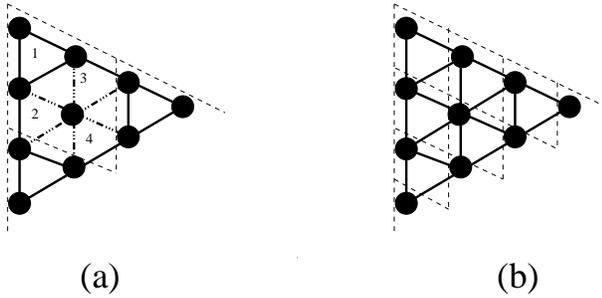}}
\vspace{0.4cm}
\caption{The $J$--$J'$ model is illustrated in diagram (a), where 
the bonds of strength $J$ between Kagom\'e lattice sites are 
indicated by the thick solid lines and the non-Kagom\'e bonds 
of strength $J'$ on the underlying triangular lattice sites 
are indicated by the ``broken'' lines. The triangular lattice
Heisenberg antiferromagnet (TAF) is illustrated in diagram (b), 
and it is noted that the two models are equivalent when $J=J'$.
The quadrilateral unit cells for both cases are also illustrated.
The $J$--$J'$ model contains four sites per unit cell, whereas 
the TAF has only one site per unit cell.}
\label{fig1}
\end{figure}

In this article we wish to apply the coupled cluster method (CCM) 
to a model which interpolates between the spin-half TAF and spin-half 
KAF models, henceforth termed the $J$--$J'$ model 
(illustrated in Fig. \ref{fig1}). 
The Hamiltonian is given by
\begin{equation}
H = J \sum_{\langle i , j \rangle} {\bf s}_i \cdot {\bf s}_j
+ J' \sum_{\{ i , k \}} {\bf s}_i \cdot {\bf s}_k ~~ ,
\label{eq1}
\end{equation}
where $\langle i , j \rangle$ runs over all nearest-neighbour (n.n.)
bonds on the Kagom\'e lattice, and $\{ i , k \}$ runs over all 
n.n. bonds which connect the Kagom\'e lattice sites to those 
other sites on an underlying triangular lattice. Note that
each bond is counted once and once only. We explicitly set 
$J=1$ throughout this paper, and we note that at 
$J'=1$ we thus have the TAF and at $J'=0$ we have the KAF.

We now briefly describe the general CCM formalism, although for 
further details the interested reader is referred to Refs. \cite{taf5,ccm1}. 
The exact ket and bra ground-state energy 
eigenvectors, $|\Psi\rangle$ and $\langle\tilde{\Psi}|$, of a 
general many-body system described by a Hamiltonian $H$, 
\begin{equation} 
H |\Psi\rangle = E_g |\Psi\rangle
\;; 
\;\;\;  
\langle\tilde{\Psi}| H = E_g \langle\tilde{\Psi}| 
\;, 
\label{ccm_eq1} 
\end{equation} 
are parametrised within the single-reference CCM as follows:   
\begin{eqnarray} 
|\Psi\rangle = {\rm e}^S |\Phi\rangle \; &;&  
\;\;\; S=\sum_{I \neq 0} {\cal S}_I C_I^{+}  \nonumber \; , \\ 
\langle\tilde{\Psi}| = \langle\Phi| \tilde{S} {\rm e}^{-S} \; &;& 
\;\;\; \tilde{S} =1 + \sum_{I \neq 0} \tilde{{\cal S}}_I C_I^{-} \; .  
\label{ccm_eq2} 
\end{eqnarray} 
The single model or reference state $|\Phi\rangle$ is required to have the 
property of being a cyclic vector with respect to two well-defined Abelian 
subalgebras of {\it multi-configurational} creation operators $\{C_I^{+}\}$ 
and their Hermitian-adjoint destruction counterparts $\{ C_I^{-} \equiv 
(C_I^{+})^\dagger \}$. Thus, $|\Phi\rangle$ plays the role of a vacuum 
state with respect to a suitable set of (mutually commuting) many-body 
creation operators $\{C_I^{+}\}$. Note that $C_I^{-} |\Phi\rangle = 0$, 
$\forall ~ I \neq 0$, and that $C_0^{-} \equiv 1$, the identity operator. 
These operators are furthermore complete in the many-body Hilbert 
(or Fock) space.
Also, the {\it correlation operator} $S$ is decomposed entirely in terms 
of these creation operators $\{C_I^{+}\}$, which, when acting on the 
model state ($\{C_I^{+}|\Phi\rangle \}$), create excitations from it.
We note that although the manifest Hermiticity, 
($\langle \tilde{\Psi}|^\dagger = |\Psi\rangle/\langle\Psi|\Psi\rangle$), 
is lost, the normalisation conditions 
$ \langle \tilde{\Psi} | \Psi\rangle
= \langle \Phi | \Psi\rangle 
= \langle \Phi | \Phi \rangle \equiv 1$ are explicitly 
imposed. The {\it correlation coefficients} $\{ {\cal S}_I, \tilde{{\cal S}}_I \}$ 
are regarded as being independent variables, and the full set 
$\{ {\cal S}_I, \tilde{{\cal S}}_I \}$ thus provides a complete 
description of the ground state. For instance, an arbitrary 
operator $A$ will have a ground-state expectation value given as, 
\begin{equation} 
\bar{A}
\equiv \langle\tilde{\Psi}\vert A \vert\Psi\rangle
=\langle\Phi | \tilde{S} {\rm e}^{-S} A {\rm e}^S | \Phi\rangle
=\bar{A}\left( \{ {\cal S}_I,\tilde{{\cal S}}_I \} \right) 
\; .
\label{ccm_eq6}
\end{equation} 

We note that the exponentiated form of the ground-state CCM 
parametrisation of Eq. (\ref{ccm_eq2}) ensures the correct counting of 
the {\it independent} and excited correlated 
many-body clusters with respect to $|\Phi\rangle$ which are present 
in the exact ground state $|\Psi\rangle$. It also ensures the 
exact incorporation of the Goldstone linked-cluster theorem, 
which itself guarantees the size-extensivity of all relevant 
extensive physical quantities. 

The determination of the correlation coefficients $\{ {\cal S}_I, \tilde{{\cal S}}_I \}$ 
is achieved by taking appropriate projections onto the ground-state 
Schr\"odinger equations of Eq. (\ref{ccm_eq1}). Equivalently, they may be 
determined variationally by requiring the ground-state energy expectation 
functional $\bar{H} ( \{ {\cal S}_I, \tilde{{\cal S}}_I\} )$, defined as in Eq. (\ref{ccm_eq6}), 
to be stationary with respect to variations in each of the (independent) 
variables of the full set. We thereby easily derive the following coupled 
set of equations, 
\begin{eqnarray} 
\delta{\bar{H}} / \delta{\tilde{{\cal S}}_I} =0 & \Rightarrow &   
\langle\Phi|C_I^{-} {\rm e}^{-S} H {\rm e}^S|\Phi\rangle = 0 ,  \;\; I \neq 0 
\;\; ; \label{ccm_eq7} \\ 
\delta{\bar{H}} / \delta{{\cal S}_I} =0 & \Rightarrow & 
\langle\Phi|\tilde{S} {\rm e}^{-S} [H,C_I^{+}] {\rm e}^S|\Phi\rangle 
= 0 , \;\; I \neq 0 \; . \label{ccm_eq8}
\end{eqnarray}  
Equation (\ref{ccm_eq7}) also shows that the ground-state energy at the stationary 
point has the simple form 
\begin{equation} 
E_g = E_g ( \{{\cal S}_I\} ) = \langle\Phi| {\rm e}^{-S} H {\rm e}^S|\Phi\rangle
\;\; . 
\label{ccm_eq9}
\end{equation}  
It is important to realize that this (bi-)variational formulation 
does {\it not} lead to an upper bound for $E_g$ when the summations for 
$S$ and $\tilde{S}$ in Eq. (\ref{ccm_eq2}) are truncated, due to the lack of 
exact Hermiticity when such approximations are made. However, one can prove  
that the important Hellmann-Feynman theorem {\it is} preserved in all 
such approximations. 

In the case of spin-lattice problems of the type considered
here, the operators $C_I^+$ become products of spin-raising
operators $s_k^+$ over a set of sites $\{k\}$, with respect to
a model state $|\Phi\rangle$ in which all spins points 
``downward'' in some suitably chosen local spin axes. 
The CCM formalism is exact in the limit of inclusion of
all possible such multi-spin cluster correlations for 
$S$ and $\tilde S$, although in any real application 
this is usually impossible to achieve. It is therefore 
necessary to utilise various approximation schemes 
within $S$ and $\tilde{S}$. The three most commonly 
employed schemes previously utilised have been: 
(1) the SUB$n$ scheme, in which all correlations 
involving only $n$ or fewer spins are retained, but no
further restriction is made concerning their spatial 
separation on the lattice; (2) the SUB$n$-$m$  
sub-approximation, in which all SUB$n$ correlations 
spanning a range of no more than $m$ adjacent lattice 
sites are retained; and (3) the localised LSUB$m$ scheme, 
in which all multi-spin correlations over all distinct 
locales on the lattice defined by $m$ or fewer contiguous 
sites are retained. 

\begin{table}
\caption{CCM results for the ground-state energy per spin and sublattice 
magnetisation of the TAF and KAF models using the LSUB$m$ approximation 
with $m=\{2,3,4,5,6\}$. CCM critical values, $J_{c}'$, of the $J$--$J'$ model
(with $J=1$), which are themselves indicators of a phase transition point 
in the true system, are also given. Comparison is made in the last row
with the results of other calculations.}
\begin{center}
\begin{tabular}{|c|c|c|c|c|c|}  \hline 
        &\multicolumn{2}{|c|}{KAF}
        &\multicolumn{2}{|c|}{TAF}
        &$J$--$J'$\\ \hline  
$m$ 	&$E_g/{N_K}$   &$M^K$   
&$E_g/N$  &$M^K$   & $J_{c}'$ \\ \hline\hline
2	&$-$0.37796         &0.8065
        &$-$0.50290         &0.8578   	
	&--                 \\ \hline
3	&$-$0.39470         &0.7338
        &$-$0.51911   	    &0.8045   	
	&$-$0.683           \\ \hline
4	&$-$0.40871         &0.6415
        &$-$0.53427 	    &0.7273   	
	&$-$0.217	    \\ \hline
5	&$-$0.41392         &0.5860
        &$-$0.53869 	    &0.6958   	
	&$-$0.244           \\ \hline
6	&$-$0.41767         &0.5504
        &$-$0.54290         &0.6561   	
	&$-$0.088           \\ \hline
$\infty$ &$-$0.4252        &0.366
        &$-$0.5505 	    &0.516	
	&0.0$\pm$0.1        \\ \hline
{\it c.f.}        
        &$-$0.43\footnote{~See Refs. \cite{kaf1,kaf2}}. 
        &0.0[{\it a]}.
        &$-$0.551\footnote{~See Ref. \cite{taf2}}.
        &0.5\footnote{~See Refs. \cite{taf3,taf4}}.
        &--                 \\ \hline
\end{tabular}
\end{center}
\label{tab1}
\end{table}

For the interpolating $J$--$J'$ model described by Eq. 
(\ref{eq1}), we choose a model state $|\Phi\rangle$ in which 
the lattice is divided into three sublattices, denoted 
$\{$A,B,C$\}$. The spins on sublattice A 
are oriented along the negative {\em z}-axis, and spins on sublattices 
B and C are oriented at $+120^\circ$ and $-120^\circ$, respectively, 
with respect to the spins on sublattice A. 
Our local axes are chosen by rotating about the 
$y$-axis the spin axes on sublattices B and C by $-120^\circ$ and 
$+120^\circ$ respectively, and by leaving the spin axes on sublattice A 
unchanged. Under these canonical transformations, 
\begin{eqnarray} 
s_B^x \rightarrow -\frac{1}{2} s_B^x - \frac{\sqrt{3}}{2} s_B^z \;\;  &;& \;\; 
s_C^x \rightarrow -\frac{1}{2} s_C^x + \frac{\sqrt{3}}{2} s_C^z \;\; , 
\nonumber \\
s_B^y \rightarrow s_B^y \;\; &;& \;\; s_C^y \rightarrow s_C^y \;\; , 
\nonumber \\ 
s_B^z \rightarrow  \frac{\sqrt{3}}{2} s_B^x -\frac{1}{2} s_B^z \;\; &;& \;\; 
s_C^z \rightarrow -\frac{\sqrt{3}}{2} s_C^x -\frac{1}{2} s_C^z \; . 
\label{ccm_j_j'_1}
\end{eqnarray} 
The model state $| \Phi \rangle$ now appears mathematically to consist
purely of spins pointing downwards along the $z$-axis, and the Hamiltonian
(for $J=1$) is given in terms of these rotated local spin axes as,
 \begin{eqnarray} 
H &=&\sum_{\langle i\rightarrow j\rangle}
\Bigl\{ 
-{1\over 2} s_i^z s_j^z
+\frac{\sqrt{3}\lambda}{4}
( s_i^z s_j^+ +s_i^z s_j^- -s_i^+ 
s_j^z - 
s_i^-s_j^z ) \nonumber \\  
&+& 
\frac{\lambda}{8}
( s_i^+s_j^- + s_i^- s_j^+ ) 
-\frac{3\lambda}{8}
( s_i^+ s_j^+ + s_i^- s_j^- ) 
\Bigl\} \nonumber \\ 
+ &J'& \sum_{\{ i\rightarrow k\}}
\Bigl\{  
-{1\over 2} s_i^z s_k^z
+\frac{\sqrt{3}\lambda}{4}
( s_i^z s_k^+ +s_i^z s_k^- -s_i^+ 
s_k^z  
- 
s_i^-s_k^z ) \nonumber \\
&+& \frac{\lambda}{8}
( s_i^+s_k^- + s_i^- s_k^+ ) 
-\frac{3\lambda}{8}
( s_i^+ s_k^+ + s_i^- s_k^- ) 
\Bigl\} ~ . \label{ccm_j_j'_2} 
\end{eqnarray}
Note that $i$ and $j$ run only over the $N_K$ sites on the Kagom\'e lattice,
whereas $k$ runs over those non-Kagom\'e sites on the (underlying) triangular 
lattice. $N$ indicates the total number of triangular-lattice sites, 
and each bond is counted once and once only. 
The symbol $\rightarrow$ indicates an explicit {\it bond directionality} 
in the Hamiltonian given by Eq. (\ref{ccm_j_j'_2}),
namely, the {\it three} directed nearest-neighbour bonds included  
in Eq. (\ref{ccm_j_j'_2}) point from sublattice sites 
A to B, B to C, and C to A for both types of bond.
We now perform high-order LSUB$m$ calculations for 
this model via a computational procedure for the Hamiltonian 
of Eq. (\ref{ccm_j_j'_2}). The interested reader is referred 
to Ref. \cite{taf5} for a full account of how such high-order 
CCM techniques are applied to lattice quantum spin systems. 

\begin{figure}
\epsfxsize=6.7cm
\centerline{\epsffile{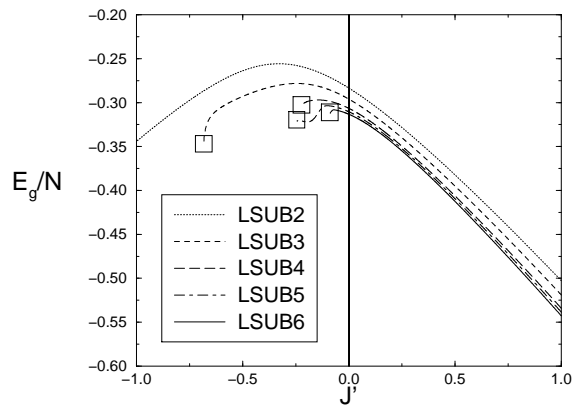}}
\vspace{0.4cm}
\caption{CCM results for the ground-state energy per spin of the 
$J$--$J'$ model (with $J=1$) 
using the LSUB$m$ approximation with $m=\{2,3,4,5,6\}$. 
The boxes indicate the CCM critical points, $J'_c$, and 
a simple extrapolation in the limit $m \rightarrow \infty$ 
implies that $J'_c=0.0\pm0.1$.}
\label{fig2}
\end{figure}

We note that for the CCM treatment of the $J$--$J'$ model presented
here the unit cell contains four lattice sites (see Fig. \ref{fig1}).  
By contrast, previous calculations \cite{taf5} for the TAF 
used a unit cell containing only a single site per unit 
cell. Hence, the $J$--$J'$ model has many more ``fundamental''
configurations than the TAF model at equivalent levels of
approximation. However, we find that those configurations 
which are {\it not} equivalent for the $J$--$J'$ model but 
{\it are} equivalent for the TAF have CCM 
correlation coefficients $\{ {\cal S}_I, \tilde {\cal S}_I \}$ 
which become equal at the TAF point, $J'=1$. Hence, the CCM 
naturally and without bias reflects the extra amount of 
symmetry of the $J$--$J'$ model at this one particular point.
This is an excellent indicator of the validity of the CCM treatment
of this model. The results for the $J$--$J'$ model at $J'=1$  
thus also exactly agree with those of a CCM previous treatment 
of the TAF. Our approach is now to ``track'' this solution for 
decreasing values of $J'$ until we reach a {\it critical} 
value of $J'_c$ at which the solution to the CCM equations 
breaks down. This is associated with a phase transition in 
the real system \cite{taf5}, and results for $J'_c$ for this 
model are presented in Table \ref{tab1}. A simple ``heuristic''
extrapolation of these results gives a value of $J'_c=0.0\pm0.1$ 
for the position of this phase transition point. This result
indicates that the classical three-sublattice N\'eel-like
order, of which about 50$\%$ remains for the TAF, completely
disappears at a point very near to the KAF point ($J'=0$).

The results for the ground-state energy are shown in Fig. \ref{fig2}
and in Table \ref{tab1}. These results are seen to be highly 
converged with respect to each other over the whole of the 
region $0 \le J' \le 1$. A simple heuristic extrapolation
may be attempted for these results for varying $J'$ by plotting 
LSUB$m$ results for $m=\{3,4,5,6\}$ against $1/m^2$ and performing 
a linear extrapolation of this data as was done previously \cite{taf5}
for the TAF only. 
These results are given in Table \ref{tab1} for the KAF and TAF 
models. We believe that the extrapolated results are among the 
most accurate results for the ground-state energies of the TAF 
and KAF ever found. 

\begin{figure}
\epsfxsize=7cm
\centerline{\epsffile{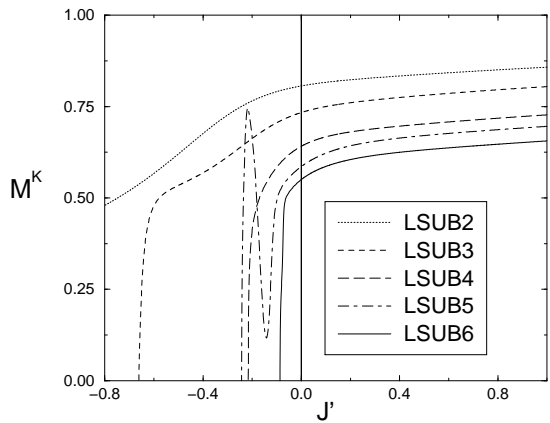}}
\vspace{0.4cm}
\caption{CCM results for the sublattice magnetisation of the $J$--$J'$
model (with $J=1$) using the LSUB$m$ approximation with $m=\{2,3,4,5,6\}$.
Note again that a simple extrapolation of LSUB$m$ critical points 
in the limit $m \rightarrow \infty$ implies that $J'_c=0.0\pm0.1$.}
\label{fig3}
\end{figure}

We now wish to consider how much of the original classical 
ordering of the model state remains for the quantum system.
Previous calculations for the TAF \cite{taf5} took the average 
value of $s^z_i$ (after rotation of the local
spin axes) where $i$ runs over {\it all} 
lattice sites. However, if one does this one also
includes non-Kagom\'e lattice sites, and when $J'=0$ the
spins on this site would be effectively ``frozen'' to 
the original direction of the model state at these sites. 
Hence, we believe that the correct order parameter for this model
is the average value of $s^z_{k}$ (again after rotation of the local
spin axes) where $k$ runs only over the $N_K$ Kagom\'e lattice 
sites. We may thus write this as, 
\begin{equation}
M^K = - \frac 2{N_K} \sum_{k=1}^{N_K} s_{k}^z ~~ .
\label{ccm_j_j'_3}
\end{equation}
The results for $M^K$ are presented in Fig. \ref{fig3} and 
in Table \ref{tab1}. Again, we extrapolate these results 
for the KAF by plotting LSUB$m$ results for $m=\{3,4,5,6\}$ 
against $1/m$ and performing a linear extrapolation of 
this data, as was done previously \cite{taf5}
for the TAF. The extrapolated result for the KAF point probably 
lies too high. However, the LSUB6 result
goes to zero very close to the KAF point, and so CCM results
are fully consistent with the hypothesis that, unlike the TAF, 
the ground state of the KAF does not contain any N\'eel ordering.

It has been shown in this article that the CCM may be used to 
provide highly accurate results for the ground-state energy 
of the $J$--$J'$ model (with $J=1$) which interpolates between the TAF and 
KAF models. Indeed, the extrapolated results for the ground-state energy 
for the KAF of $E_g/N_K$=$-$0.4252 and for the TAF of $E_g/N$=$-$0.5505
are among the most accurate yet determined for these models. 
Furthermore, the amount of classical ordering (evaluated on the 
Kagom\'e lattice sites only) yields results which are fully
consistent with the hypothesis that the KAF is fully disordered.
CCM critical points also reinforce the conjecture that the 
classically ordered phase evident for the TAF breaks down very 
near to the KAF point.

\vspace{2cm}

\end{document}